\documentclass[manuscript,screen,nonacm]{acmart}

\AtBeginDocument{%
  \providecommand\BibTeX{{%
    \normalfont B\kern-0.5em{\scshape i\kern-0.25em b}\kern-0.8em\TeX}}}

\setcopyright{none}
\acmConference[Preprint]{arXiv preprint}{2026}{}
\acmYear{2026}
\acmISBN{}
\acmDOI{}
\authorsaddresses{Author's Contact Information: Joonhyung Bae, Korea Advanced Institute of Science and Technology (KAIST), Daejeon, South Korea; jh.bae@kaist.ac.kr; ORCID 0000-0001-5933-4302.}

\begin{document}

\title[MinhwaNet: Korean Folk Painting Object Grounding]{MinhwaNet: Faithful but Insufficient Object Grounding in Korean Folk Painting}

\author{Joonhyung Bae}
\orcid{0000-0001-5933-4302}
\email{jh.bae@kaist.ac.kr}
\affiliation{%
  \institution{Korea Advanced Institute of Science and Technology (KAIST)}
  \city{Daejeon}
  \country{South Korea}
}

\renewcommand{\shortauthors}{J. Bae}

\begin{abstract}
Korean folk painting (minhwa) is built from a small vocabulary of auspicious symbols, a tiger for protection, a pair of birds for marital harmony, a peony for wealth, that recur across many of its painted genres. This suggests an obvious computational approach, identify which symbols appear in a painting and read the genre from the inventory. Working with a public corpus that pairs whole paintings, eight-field bilingual curatorial captions, and a separate set of expert object crops, we find that this approach does not work. A model given only a list of which symbols a painting contains predicts the genre far worse than a model that fuses the image with the curatorial text, and forcing the genre representation to be object-grounded actively hurts accuracy. The visual evidence on which the genre prediction rests is nonetheless localized and inspectable. A leakage-safe object evidence map projected from a part-level detector is spatially faithful to where curators isolated symbolic objects and to a patch-based surrogate's own gradient saliency. We name this configuration a faithful-but-insufficient dissociation. The part-level explanation is honest about what the part-level model sees, yet the genre target turns on how symbols are arranged rather than on which ones appear. The same lens separates a content label that survives transfer to held-out source institutions, genre, from a style label that does not, era, a prediction we confirm on two further labels in the corpus. We release the multimodal system, a worked-example reading of one painting's evidence map against its catalogue, and a set of evaluation cautions that recur in long-tailed heritage collections.
\end{abstract}

\ccsdesc[500]{Applied computing~Fine arts}
\ccsdesc[500]{Computing methodologies~Object recognition}
\ccsdesc[300]{Information systems~Multimedia and multimodal retrieval}
\ccsdesc[300]{Computing methodologies~Supervised learning by classification}

\keywords{Korean folk painting, cultural heritage, multimodal classification, model interpretability}

\maketitle

\section{Introduction}\label{sec:intro}

Korean folk painting, known as minhwa, is built from a small lexicon of auspicious symbols. A tiger wards off misfortune, a pair of birds promises marital harmony, a peony stands for wealth. These symbols recur across many of its painted genres, so a tiger appears not only in tiger paintings but also in narrative scenes, and a peony can flower in a bird-and-flower work, in a flower painting, or in a still life. This recurrence raises an obvious computational question. If the same symbols appear in many genres, can a model read the genre from the visible inventory of symbols, or is genre fixed instead by how symbols are arranged, painted, and brought into one composition? The question, as we will show, has a definite negative answer that recasts the role of object grounding in cultural-heritage computing.

Minhwa was painted largely by anonymous artisans for everyday and ceremonial use, decorating homes and folding screens with auspicious imagery rather than for the literati canon~\cite{ref:minhwa-handbook}. Curators organize its subjects into a system of hwamok, thematic genres that include bird-and-flower painting, character paintings of Confucian virtues in stylized letters, books-and-scholars still lifes, landscapes, and animal and Daoist-figure paintings. Public digitization has produced large, expertly described collections, yet computational study of minhwa remains limited compared with Western art and existing work typically treats the image or the text in isolation~\cite{ref:artsurvey,ref:nca-painting,ref:korean-clustering,ref:choson-objects}.

The public corpus we use is well suited to the question we pose. Each whole painting carries a curatorial genre label, eight Korean and English caption fields describing the work, and metadata on era, source institution, and material. Alongside the whole paintings the corpus releases a separate set of expert object crops, each cut from a source painting and described by its own localized captions. This pairing of whole-artwork records with part-level object records lets us study the part-to-whole relationship between depicted symbols and genre directly, without ad-hoc bounding-box annotation.

We build MinhwaNet, a multimodal system, and use it to answer three questions of interest to computational cultural heritage. When does multimodal fusion of image and curatorial text help genre classification, and is the gain robust to the collection-specific style signatures that pervade heritage data. Can the symbolic objects that drive genre be grounded from the expert crops alone, without bounding-box supervision, given that each symbol typically occupies a small region of the canvas. Does grounding the genre representation in those objects improve genre accuracy, or is their value interpretive. Two of the answers are about what one would expect. Fusion reliably helps a content label like genre, with the gain concentrated on the long tail, and a patch-level multiple-instance model significantly outperforms a global image embedding at recognizing symbolic objects~\cite{ref:milattn,ref:cam,ref:asl}. The third answer is not. Forcing the genre representation to be object-grounded does not raise genre accuracy and a model given only an object inventory predicts the genre poorly. The visual evidence on which the genre prediction rests is nonetheless localized and inspectable, both functionally against deletion baselines including the model's own gradient saliency and spatially against expert crop locations. We name this configuration a faithful-but-insufficient dissociation, where the part-level explanation is honest about what the part-level model sees but the genre target is not reducible to it.

The same lens organizes a second observation. Fusing image and text helps a content label whose attributes the captions describe, like genre, and does not help a style label that the captions ignore, like era. The same separation survives under transfer to held-out source institutions. We test this content-versus-style prediction on two further labels in the corpus, theme and drawing technique, and find that it holds.

Our contributions are diagnostic findings established with standard components rather than new architectures. We construct a multimodal genre model, a part-level object grounder, and a leakage-safe whole-artwork object evidence map. With them we characterize when curatorial text helps a visual heritage classifier and when it does not, with cross-source evaluation discriminating fusion designs that are indistinguishable in distribution. We show that the object evidence map is faithful to a patch-based surrogate predictor against random, shuffled, patch-norm, and gradient-saliency baselines and that it agrees spatially with expert crop locations. We demonstrate the faithful-but-insufficient dissociation, where a part-level attribution can be honest yet the target it explains is defined by arrangement and rendering rather than by the inventory it lists, a caution for explanation methods on any compositional category. We release code, trained weights, fold assignments, and derived count matrices.

\section{Related work}\label{sec:related}

\subsection{Computational analysis of paintings}

Deep learning has been applied widely to the analysis of paintings, predicting genre, style, period, and artist from the image alone. Large benchmarks have driven this progress, with WikiArt-style recognition supplying tens of thousands of labeled Western paintings and the OmniArt benchmark adding artist, type, and timeframe annotations at scale~\cite{ref:wikiart,ref:omniart}. Surveys of the area report steady gains from convolutional and transformer backbones, usually a strong pretrained encoder with a lightweight classification head, alongside a recurring difficulty, the heavy class imbalance of curated collections where a few canonical categories dominate and a long tail of rare ones is modeled poorly~\cite{ref:artsurvey,ref:nca-painting}. Our setting shares this long-tailed structure but differs in subject, since minhwa genres are organized around recurring symbolic themes rather than the movements and schools that label Western art, so the categories overlap in their visual content.

\subsection{East Asian and Korean painting}

Computational study of East Asian painting is comparatively recent and works with fewer labeled resources than the Western canon. Work on Chinese painting has addressed style and dynasty classification and the segmentation of ink-wash imagery, where conventions such as expansive empty space and calligraphic line differ from those that pretrained Western-art models expect~\cite{ref:chinese-style}. A smaller body of work on Korean painting has clustered twentieth-century works from combined visual and textual features~\cite{ref:korean-clustering}. Korean folk painting has received little computational attention despite its large public digitization, and existing studies treat either the image or the text in isolation. We use a corpus that pairs whole paintings with expert object crops and with bilingual field-structured captions, which lets us study the image and the curatorial text jointly rather than separately.

\subsection{Multimodal fusion for cultural heritage}

Modern multimodal classification rests on contrastive image-text pretraining and on transformer encoders that produce contextual token representations~\cite{ref:clip,ref:vit,ref:transformer}. Surveys organize image-text fusion into early fusion that joins features, late fusion that combines decisions, and attention-based schemes that let one modality attend over the other, of which the cross-attention design we use is one instance~\cite{ref:mmsurvey,ref:vilbert}. In cultural heritage, image-text pairs have been used to assign metadata, retrieve across modalities, and build semantic understanding of artworks from paired descriptions~\cite{ref:chmeta,ref:culti,ref:semart}. Heritage captions are often field structured rather than free form, as in the eight curatorial fields of our corpus, which we keep as separate tokens instead of flattening into one string. These systems generally assume that the text helps, whereas we ask where curatorial text helps a visual genre classifier and where it does not, and we find the answer turns on whether the target is a content label or a style label.

\subsection{Weakly-supervised localization and faithfulness}

Recognizing which symbolic objects a painting contains, without box supervision, connects to multiple-instance learning and to weakly supervised localization, where image-level labels train a model whose intermediate activations localize the responsible regions~\cite{ref:milattn,ref:cam,ref:gradcam}. Small symbolic objects occupy a fraction of the canvas, so a single global image embedding tends to underrepresent them and patch-level evidence becomes informative. Multi-label recognition with many rare positives is commonly stabilized by asymmetric losses that down-weight easy negatives~\cite{ref:asl}, and iconography recognition in paintings is the closest art-domain instance of the task~\cite{ref:artdl}. Because a localization map can look plausible while not reflecting the model, its evaluation requires faithfulness measures rather than visual inspection, for which deletion and insertion metrics and dedicated localization-evaluation protocols are standard~\cite{ref:rise,ref:wsoleval}. We adopt the deletion and insertion methodology and test the object evidence map against an explicit genre predictor rather than reading the map by eye.

\subsection{Dataset confounds and transfer}

A growing literature shows that classifiers exploit spurious correlations, so within-dataset accuracy overstates what a model has learned and often fails to transfer to new sources~\cite{ref:shortcut,ref:leakage,ref:reforms}. Heritage collections are especially prone to this, since holdings concentrate in a few institutions whose acquisition and digitization practices imprint a collection-specific signature on the images~\cite{ref:collections}. We bring this lens to era attribution, where a within-collection signal that looks temporal turns out to track the institution, and we adopt leakage-safe practice throughout, masking templated label text and splitting by source. For the long-tailed genre distribution we use the standard remedies of reweighting and class-balanced losses~\cite{ref:cb,ref:focal}.

\section{Dataset}\label{sec:data}

We use a public collection of 8,206 Korean folk paintings with bilingual eight-field captions and curatorial metadata, distributed under a national data policy. Each whole-painting record carries a genre label, one of sixteen hwamok, and most carry a production era among late Joseon, the colonial period, and the post-liberation period. The eight caption fields are overview, composition, production, symbolism, keyword, integrated summary, and two object descriptions, each in Korean and English. For genre we use the 3,411 whole paintings that carry a genre label. Table~\ref{tab:taxonomy} lists the sixteen hwamok with their counts. The vocabulary mixes subject-defined classes such as bird-and-flower and landscape with classes defined by depicted function or material such as vessels-and-implements and scholar-stone, and several genres are distinguished less by a single object than by the symbolic theme that organizes the scene. The sixteen-class scheme is a curatorial construct rather than a settled art-historical canon, and boundaries between related genres are fluid, with the composite class in particular functioning partly as a residual category for mixed-subject works. The distribution is strongly long tailed, ranging from 826 paintings for the bird-and-flower genre to 40 for architecture, with the three largest genres accounting for about three fifths of the data and the rarest six each below sixty examples, so macro-averaged metrics are dominated by the long tail. One institution contributes the majority of the holdings, a source imbalance we return to as a confound.

\begin{table}[t]
\caption{The sixteen hwamok classes and their counts among the 3,411 labeled paintings.}
\label{tab:taxonomy}
\begin{tabular}{llr}
\toprule
Genre & hwamok & $n$ \\
\midrule
Bird-and-flower & hwajodo & 826 \\
Character & munjado & 683 \\
Landscape & sansuhwa & 586 \\
Fish-and-crab & eohaedo & 235 \\
Composite & honseongdo & 185 \\
Flower & hwahwedo & 183 \\
Vessels-and-implements & giyonghwa & 149 \\
Figure & inmulhwa & 116 \\
Narrative & seolhwahwa & 86 \\
Auspicious-animal & yeongsuhwa & 71 \\
Fruit-and-vegetable & sogwado & 57 \\
Animal & chuksudo & 49 \\
Scholar-stone & suseokdo & 49 \\
Grass-and-insect & chochungdo & 49 \\
Daoist-figure & doseokhwa & 47 \\
Architecture & oguhwa & 40 \\
\bottomrule
\end{tabular}
\end{table}

The dataset also provides a detailed object-crop subset, in which curators cropped individual symbolic elements and described each crop with localized captions. We use 4,795 object crops, each linked to its source painting by a shared identifier. From the structured object and symbolism fields we derive two multi-label target sets, a stricter object vocabulary of 60 labels and a broader motif vocabulary of 80 labels, with on average 0.66 and 2.25 labels per crop. These targets are derived from metadata text and exported only as aggregate count matrices, never as raw captions. Each crop inherits the genre fold of its source painting, so crops from a held-out painting are excluded from the corresponding training fold.

We control for label leakage in the genre text. The templated overview field names and defines the genre, so we drop it and mask all sixteen genre names and the classification template phrases across the remaining fields. After masking, zero genre-name tokens remain across the 3,411 paintings. A classifier given only the masking pattern, the per-field counts of the mask token, recovers genre at 0.071 macro-F1, essentially the 0.062 uniform chance, and field length alone gives 0.132, whereas a content bag-of-words on the masked text reaches 0.806 with iconographic content nouns rather than label phrases as its most discriminative tokens. The residual text signal used for genre is therefore genuine described content rather than a stated label. The dataset cannot be redistributed, so we release derived artifacts and code rather than raw images.

\section{Method}\label{sec:method}

\begin{figure*}[t]
\centering
\includegraphics[width=\textwidth]{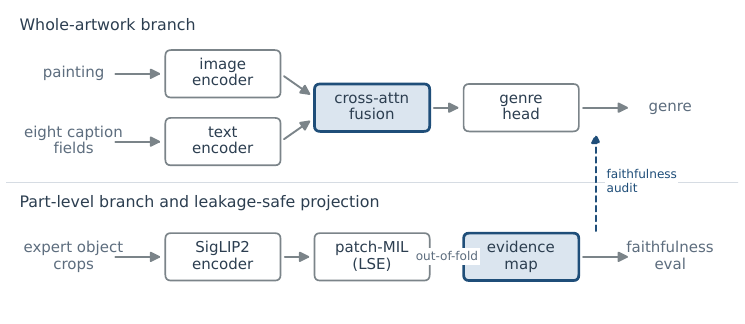}
\caption{System overview. The whole-artwork branch (top) fuses image and text into a genre prediction. The part-level branch (bottom) projects an object detector onto whole paintings out of fold to produce a leakage-safe evidence map that the faithfulness analysis audits.}
\label{fig:overview}
\end{figure*}

Figure~\ref{fig:overview} summarizes the two branches of the system, the whole-artwork multimodal branch that predicts genre and the part-level branch that grounds symbolic objects, connected by a leakage-safe projection of the part-level detector onto whole paintings.

\subsection{Multimodal genre model}

MinhwaNet encodes the painting and its caption separately and combines them. A frozen vision transformer produces the image class token, and a frozen text encoder produces one contextual embedding per caption field, so long captions are not truncated and the field structure is preserved~\cite{ref:clip,ref:vit}. Within this system we compare four fusion configurations, image only, text only, concatenation, and a cross-attention model that uses the image class token as a query over the eight caption-field tokens. Writing the image class token as $\mathbf{q}$ and the eight caption-field embeddings as $\mathbf{t}_1,\dots,\mathbf{t}_8$, the cross-attention fusion projects a query, keys, and values and returns an attended caption summary
\begin{equation}
\mathbf{c} = \sum_{j=1}^{8} \alpha_j\, W_v \mathbf{t}_j, \qquad
\alpha_j = \operatorname*{softmax}_j \frac{(W_q \mathbf{q})^\top (W_k \mathbf{t}_j)}{\sqrt{d}},
\end{equation}
which is concatenated with the image token before the head. Concatenation instead joins the image token with the mean of the field embeddings without the learned weighting $\alpha_j$. Every configuration uses the same head, per-feature standardization followed by a two-layer perceptron with dropout.

\subsection{Weakly-supervised object grounding}

We recognize symbolic objects and motifs from the expert crops as a multi-label task. Each crop is encoded by a frozen SigLIP2 vision transformer, whose dense features suit localizing small symbolic regions~\cite{ref:siglip2}. A global baseline classifies each crop from its pooled image token. Because object evidence is localized, we add a patch-level multiple-instance model. A layer-normalized linear head predicts a logit $z_{c,i}$ for class $c$ at each of the $P=196$ patch tokens of the fourteen by fourteen grid, and the crop-level patch score is a temperature-scaled log-sum-exp smooth maximum over patches
\begin{equation}
s_c = \tau \log \frac{1}{P} \sum_{i=1}^{P} \exp\!\big(z_{c,i}/\tau\big),
\end{equation}
with temperature $\tau=0.5$, which approaches the patch maximum as $\tau$ shrinks and a mean as it grows. The patch score is combined with the global-token logit $g_c$ through a learned per-class weight $\lambda_c$ into the prediction $\sigma(\lambda_c s_c + (1-\lambda_c) g_c)$~\cite{ref:milattn,ref:cam}. The head is trained with an asymmetric loss that down-weights easy negatives, with negative focusing exponent four and probability margin 0.05~\cite{ref:asl}.

\subsection{Whole-artwork object evidence map}

To connect the part level to genre we project the trained crop detector onto whole paintings. For each genre fold the detector is trained only on crops whose source painting lies outside that fold, then applied to the held-out paintings, which yields a leakage-safe object evidence map for every painting. The map has two forms, a vector of object probabilities and a grid of per-patch object scores aligned to the painting patches.

\subsection{Training and evaluation}

Encoders are frozen and only the head and fusion parameters are trained, which isolates the fusion contribution. The whole-artwork image encoder is a CLIP vision transformer of base size at patch sixteen, and the part-level encoder is SigLIP2 of base size at patch sixteen. We optimize a class-weighted cross-entropy with AdamW at learning rate $10^{-3}$, select the epoch by macro-F1 on a validation split drawn only from the training folds, and evaluate once on the test fold under a fixed seed. We report macro-F1 as the primary metric, with average precision for the threshold-free part-level task, as five-fold mean and standard deviation, and attach paired significance to every reported comparison using a paired bootstrap with 10{,}000 resamples over the per-item predictions pooled across folds. Multi-label F1 uses a fixed 0.5 threshold, and patch deletion replaces a patch token with the per-fold mean. Protocol P1 is a label-stratified five-fold split over the 3,411 labeled paintings, P2 restricts both training and testing to the single dominant collection to remove the source confound, and P3 trains on the dominant collection and tests on the remainder to measure cross-source transfer. Part-level folds are defined by the source painting, so crops and whole paintings never cross the train and test boundary. The released artifacts include the fold assignments, seeds, and per-fold weights, so every number is reproducible from features even where the raw images cannot be redistributed.

\section{Results}\label{sec:results}

\subsection{When does multimodal fusion help genre}

Table~\ref{tab:genre} reports genre under P1. Multimodal fusion exceeds both single modalities, with concatenation at 0.889 macro-F1 and cross-attention at 0.886 against 0.837 for text and 0.773 for image. A paired bootstrap confirms the gap, with concatenation beating text by 0.053 (95\% CI [0.036, 0.071], $p<10^{-4}$) and image by 0.116, while the fusion variants are mutually indistinguishable. The gain concentrates on rare and visually confusable genres, where averaged over genres with fewer than sixty examples the fusion gain is 0.053 against 0.024 for genres with at least two hundred. The advantage is not a source artifact, since within the single dominant collection under P2 the fusion ordering is preserved, with concatenation at 0.816 and cross-attention at 0.793 against 0.770 for text and 0.701 for image. Concatenation is thus the strongest fusion both in distribution and within a single collection, and the cross-attention variant is preferred only for its cross-source robustness (Section~\ref{sec:cross}). Text alone outperforms image alone because the described content names the depicted objects that define a genre even after the genre label is masked. The fusion gain is therefore available when a structured caption already exists at prediction time, as in metadata enrichment or consistency checking of described collections, while for an undescribed painting the image-only result applies.

\begin{table}[t]
\caption{Genre classification (sixteen hwamok) under P1, five-fold mean and standard deviation.}
\label{tab:genre}
\begin{tabular}{lcc}
\toprule
Model & macro-F1 & accuracy \\
\midrule
Image only & 0.773 $\pm$ 0.029 & 0.865 \\
Text only (genre masked) & 0.837 $\pm$ 0.015 & 0.898 \\
Concatenation & 0.889 $\pm$ 0.021 & 0.934 \\
Cross-attention & 0.886 $\pm$ 0.016 & 0.931 \\
\bottomrule
\end{tabular}
\end{table}

For reference, a CLIP ViT-B/16 zero-shot classifier over English prompts of the genre names reaches 0.305 macro-F1, far below the trained image-only head at 0.773, indicating that minhwa genre recognition requires a task-specific layer over the pretrained features rather than direct prompt-based recognition. A field-wise text-only ablation, removing each of the eight caption fields in turn and retraining the text-only head, shows that the two object-description fields carry the largest share of the text signal, with removing object1 dropping text macro-F1 by 0.042 and object2 by 0.028, while removing the production or overview field drops it by less than 0.01. The text signal is therefore concentrated in descriptions of depicted objects rather than in template or production-method language, consistent with genre being a content-defined label.

We use frozen encoders to isolate the fusion contribution, and a budget-matched control confirms this does not understate the image modality. Adapting the vision encoder end to end with rank eight LoRA adapters ($\alpha=16$, dropout 0.1) under a trainable-parameter budget matched to the head did not exceed the frozen features, reaching 0.713 genre macro-F1 against 0.773 frozen and 0.679 era against 0.772 frozen, consistent with the strength of the pretrained features on a few thousand images.

\begin{figure}[t]
\centering
\includegraphics[width=0.7\columnwidth]{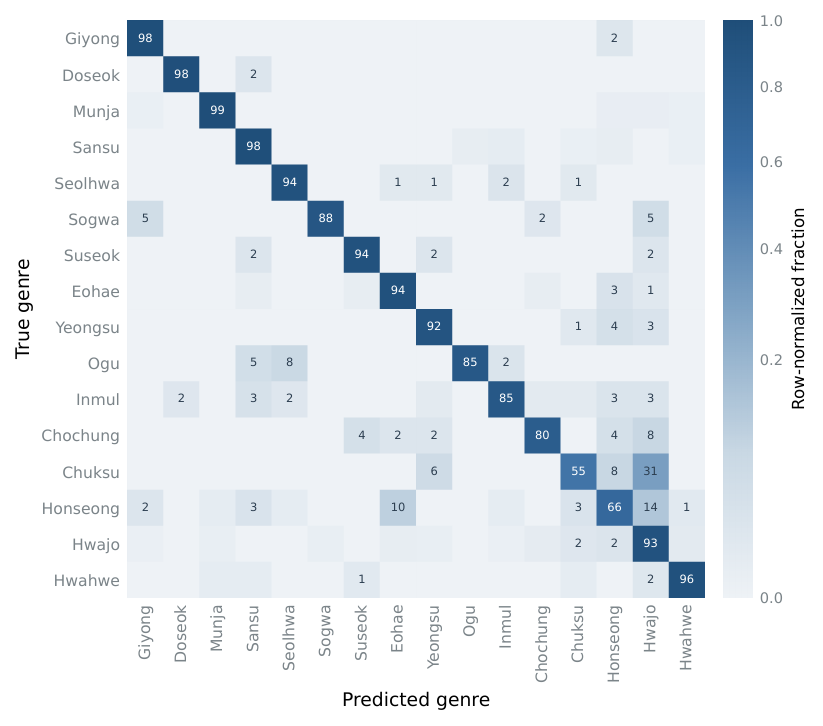}
\caption{Row-normalized confusion matrix for the cross-attention model under P1. Cells in percent.}
\label{fig:confusion}
\end{figure}

Figure~\ref{fig:confusion} shows the row-normalized confusion matrix of the cross-attention model. Most mass lies on the diagonal, and the residual confusions fall between genres with similar object content.

The off-diagonal mass is not spread uniformly but concentrates between genres that share symbolic objects. The composite genre, which by definition combines elements of others, is confused most often with bird-and-flower and with flower painting, and the two floral genres are confused with each other, since all three depict peonies, birds, and blossoming branches. The rare genres carry most of the residual error, as expected from the long tail, and their confusions land on the larger genre with the most similar object inventory rather than on an arbitrary class. This object-structured pattern, which we quantify against per-genre object profiles in Section~\ref{sec:bridge}, is what motivates asking whether the object inventory itself determines genre.

\subsection{Cross-source transfer}\label{sec:cross}

Table~\ref{tab:cross} reports transfer across collections under P3. Genre transfers and the cross-attention fusion is the most robust to the move. A paired bootstrap over the held-out collections shows that plain concatenation falls significantly below text alone ($-0.043$, 95\% CI $[-0.082, -0.004]$, $p=0.027$), because the non-transferring image features it concatenates drag the fusion below the transferable text signal, while cross-attention matches text-only robustness ($-0.009$, $p=0.58$) and significantly beats concatenation ($+0.034$, $p=0.024$). The image transfers poorly because visual style is collection specific, whereas caption content is collection invariant. Era does not transfer and collapses to the three-class chance level for every modality.

The mechanism behind the cross-attention robustness is visible in its field-attention. Under P1, the cross-attention model concentrates an average 0.85 of its attention on the keyword field, with object1 and object2 receiving only 0.07 and 0.02. Under P3, the attention shifts. Keyword falls to 0.75 and the two object-description fields rise to 0.13 and 0.07. The transferable signal across source institutions is the concrete description of depicted objects, and the cross-attention model selects those fields under transfer where plain concatenation, which averages all fields equally, cannot.

\begin{table}[t]
\caption{Cross-source transfer under P3, macro-F1. P3 is a single source-disjoint split, so no fold deviation is reported.}
\label{tab:cross}
\begin{tabular}{lcc}
\toprule
Model & Genre P3 & Era P3 \\
\midrule
Image only & 0.527 & 0.368 \\
Text only & 0.784 & 0.396 \\
Concatenation & 0.740 & 0.385 \\
Cross-attention & 0.774 & 0.384 \\
\bottomrule
\end{tabular}
\end{table}

\subsection{Probing the dichotomy on additional labels}\label{sec:dichotomy}

The content-versus-style prediction was made on two labels, genre and era. To test it further we apply the same image-only and text-only pipelines to two additional curatorial labels in the corpus, theme\_type with nine higher-level subject categories such as animals-and-plants, character, and landscape, and drawing technique with six single-style categories such as colored painting, light wash, and ink wash. The first is a content label, since the captions describe the depicted subject, and the second a style label, since the rendering technique is rarely named in the description. Table~\ref{tab:dichotomy} reports the result. For theme\_type, text-only reaches 0.797 macro-F1 under P1, close to image-only at 0.847, and transfers to held-out source institutions at 0.703 under P3, the same pattern as genre. For drawing technique, text-only collapses below image-only at 0.617 against 0.873 under P1, and fails to transfer at P3 text 0.452 against image 0.536, the same pattern as era. The four-label test holds the prediction, with the underlying mechanism being whether the curatorial text describes the target attribute.

\begin{table}[t]
\caption{Image- and text-only macro-F1 on four labels under P1 and P3. Theme and drawing use the same features with a head matched across labels.}
\label{tab:dichotomy}
\begin{tabular}{llcc cc}
\toprule
 & & \multicolumn{2}{c}{P1 macro-F1} & \multicolumn{2}{c}{P3 macro-F1} \\
\cmidrule(lr){3-4}\cmidrule(lr){5-6}
Label & type & image & text & image & text \\
\midrule
Genre (16) & content & 0.773 & 0.837 & 0.527 & 0.784 \\
Theme (9) & content & 0.847 & 0.797 & 0.632 & 0.703 \\
Drawing (6) & style & 0.873 & 0.617 & 0.536 & 0.452 \\
Era (3) & style & 0.772 & 0.580 & 0.368 & 0.396 \\
\bottomrule
\end{tabular}
\end{table}

\subsection{Weakly-supervised object grounding}

Table~\ref{tab:part} reports part-level recognition. For the stricter object vocabulary, the patch-level multiple-instance model raises micro-averaged precision from 0.450 to 0.485 over the global image token, and the motif vocabulary improves from 0.292 to 0.350. A paired bootstrap over the pooled out-of-fold crops confirms that every average-precision gain is significant. The object micro-AP gain has a 95\% CI of [0.013, 0.040] ($p<10^{-4}$), object macro-AP improves at $p=0.025$, and motif micro and macro-AP both improve at $p<10^{-4}$. The thresholded macro-F1 for motif slightly regresses (0.299 to 0.286) even as the threshold-free macro-AP rises, a thresholding effect on rare and abstract motif labels under the fixed operating point rather than a ranking loss. The text rows are an upper bound induced by the target construction, since the targets derive from the caption fields, so the fair comparison is the global token against patch aggregation, which confirms that minhwa object evidence is localized.

\begin{table}[t]
\caption{Part-level multi-label recognition on expert object crops, five-fold mean and standard deviation over source-painting folds. Text rows are an upper bound induced by the target construction.}
\label{tab:part}
\begin{tabular}{llcccc}
\toprule
Target & Model & micro-AP & macro-AP & micro-F1 & macro-F1 \\
\midrule
Object (60) & Image, global token & 0.450 $\pm$ 0.024 & 0.441 $\pm$ 0.029 & 0.475 $\pm$ 0.025 & 0.356 $\pm$ 0.032 \\
Object (60) & Image, patch-MIL & 0.485 $\pm$ 0.032 & 0.468 $\pm$ 0.030 & 0.493 $\pm$ 0.017 & 0.370 $\pm$ 0.026 \\
Object (60) & Text (oracle) & 0.910 $\pm$ 0.007 & 0.890 $\pm$ 0.012 & 0.846 $\pm$ 0.009 & 0.782 $\pm$ 0.026 \\
\midrule
Motif (80) & Image, global token & 0.292 $\pm$ 0.011 & 0.295 $\pm$ 0.012 & 0.328 $\pm$ 0.003 & 0.299 $\pm$ 0.012 \\
Motif (80) & Image, patch-MIL & 0.350 $\pm$ 0.015 & 0.323 $\pm$ 0.010 & 0.376 $\pm$ 0.015 & 0.286 $\pm$ 0.016 \\
Motif (80) & Text (oracle) & 0.824 $\pm$ 0.010 & 0.776 $\pm$ 0.012 & 0.755 $\pm$ 0.011 & 0.711 $\pm$ 0.014 \\
\bottomrule
\end{tabular}
\end{table}

\subsection{Object grounding is an interpretable bridge, not a performance lever}\label{sec:bridge}

The object evidence map is faithful to the genre prediction, which we verify without human annotation against a patch-based surrogate genre predictor, a logistic head over mean-pooled painting patches, rather than the cross-attention model that consumes a single class token. The surrogate is intentionally simple so that its own gradient saliency, the upper-faithfulness reference used below, is well understood by construction. We rank each painting's patches by their aggregate object evidence and delete them in that order. Deleting evidence-ranked patches degrades the true-genre probability far faster than deleting random patches, with per-painting deletion area under the curve 0.612 against 0.783, a gap of 0.171 (95\% CI [0.164, 0.178], $p<10^{-4}$, $n=3411$). Three controls calibrate the reading. A shuffled-evidence map matches the random baseline exactly (0.783), confirming the effect comes from the object evidence and not from deletion alone. Evidence ranking beats a generic patch-norm saliency (0.693). Against the surrogate's own gradient saliency (0.302), which is the most faithful ordering by construction, the object evidence recovers about a third of the gradient's deletion effect while remaining well above random, so the map is a useful but not optimal attribution and should be read alongside a gradient saliency where the deployed model permits one rather than on its own. A second surrogate, a small two-layer perceptron over the same patch mean, replicates the direction with evidence deletion AUC 0.691 against the random baseline 0.718, confirming the faithful-but-insufficient pattern across surrogate families while showing the gap to be larger under the more linear logistic head.

\begin{figure*}[p]
\centering
\includegraphics[width=0.9\textwidth]{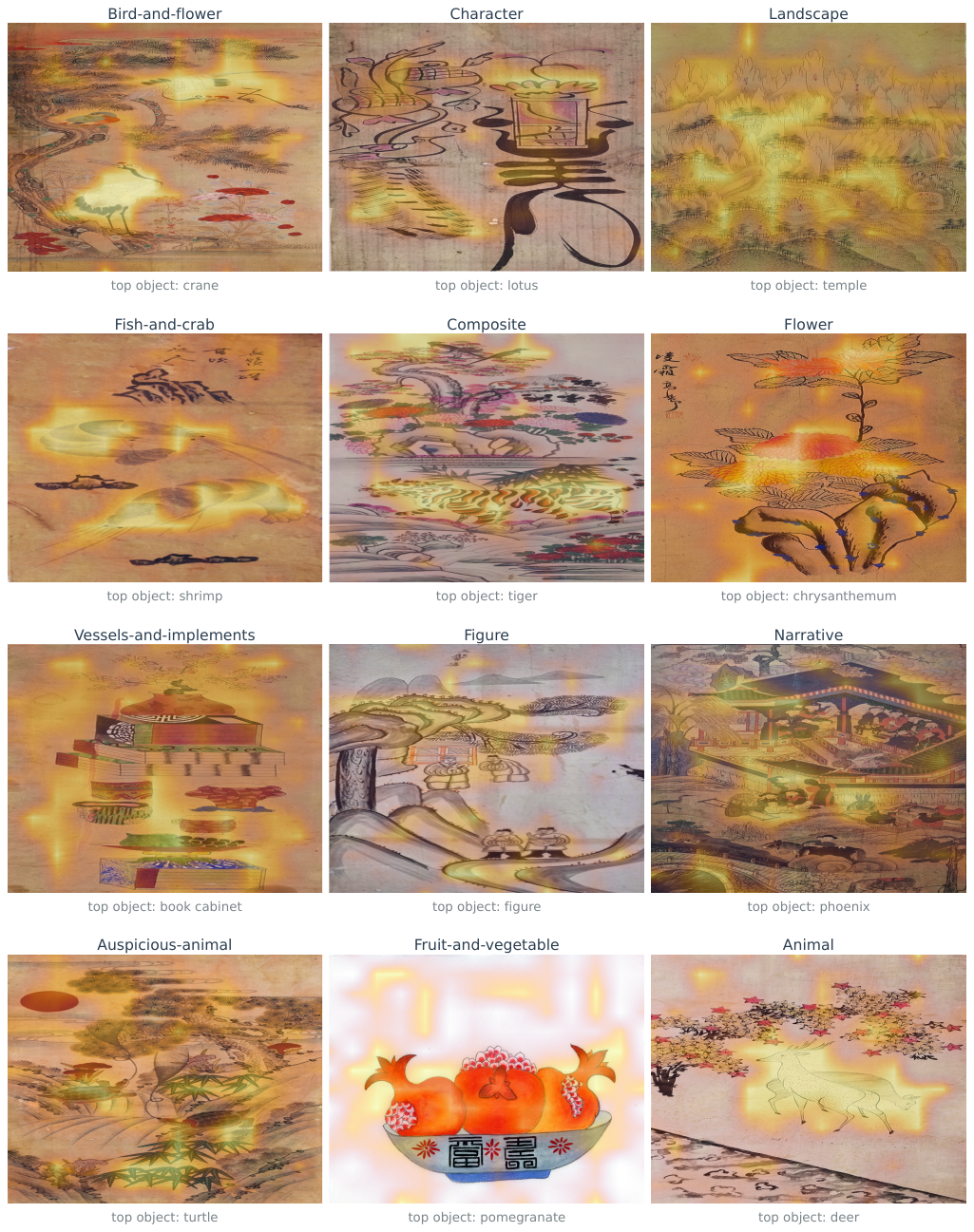}
\caption{Object evidence overlaid on twelve representative paintings, one per genre. Per-panel label gives the model's top-1 object with its patch score. Images \textcopyright\ Gahoe Minhwa Museum, National Folk Museum of Korea, Seoul Museum of History, Samcheok Municipal Museum, and private collections; all distributed under Korea Open Government License Type 1.}
\label{fig:overlay}
\end{figure*}

\begin{figure}[t]
\centering
\includegraphics[width=0.7\columnwidth]{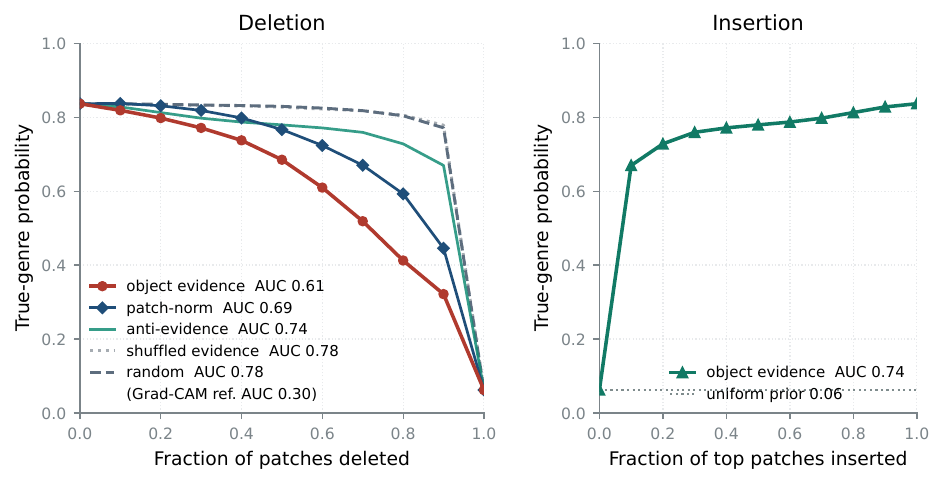}
\caption{Faithfulness of the object evidence to a patch-based surrogate genre predictor. Left, deletion under five orderings (lower curve is more faithful). The Grad-CAM AUC 0.30 in the legend is the upper-faithfulness reference. Right, insertion of the top evidence-ordered patches.}
\label{fig:faith}
\end{figure}

A spatial check complements the functional deletion test. For each expert object crop we localize its position in the source painting by ORB keypoint matching with RANSAC homography on the crop and the whole image, recovering a bounding box for 4{,}243 of the 4{,}337 crops linked to a labeled painting (97.8\%). The intersection-over-union between the top-twenty evidence-ranked patches and the crop bounding box averages 0.113 against 0.087 for a shuffled-evidence control, and 59.3\% of crops achieve an IoU above 0.1 against 31.3\% for the shuffled control. The object evidence therefore exhibits spatial agreement with where curators isolated symbolic elements, beyond the functional deletion-insertion signal. Inserting only the top tenth of evidence-ranked patches recovers the true-genre probability from the 0.06 uniform-genre prior to 0.67. Figure~\ref{fig:overlay} shows the evidence overlaid on a painting and Figure~\ref{fig:faith} the curves.

Faithfulness does not mean the objects determine genre. A logistic-regression classifier given only the artwork-level object inventory reaches 0.549 macro-F1 under P1, and the broader motif inventory reaches 0.492, both far below the multimodal genre model at 0.886 and above the majority baseline of 0.176. The inventory is derived from text and is therefore a generous proxy, an upper bound on what a purely visual inventory could supply. We cannot fully separate a genuinely relational reading of genre from extraction loss in this proxy, but because the proxy is generous the large remaining gap still indicates that a list of which symbols are present does not name the genre, which depends on how the objects are composed and rendered. The same conclusion holds when grounding is pushed into the genre representation. Table~\ref{tab:bridge} reports four ways of injecting the object evidence into a SigLIP2 genre model whose baseline reaches 0.896 macro-F1. An object-evidence expert added to the baseline sits at the baseline ($-0.002$, $p=0.69$) and object-guided pooling that weights painting patches by their object scores ties it ($+0.001$, $p=0.87$), while the two stronger objectives that force the representation to be object-grounded both lower genre accuracy significantly, distilling the evidence map into a learned patch encoder to 0.883 ($-0.013$, $p=0.020$) and two-stage object pretraining to 0.884 ($-0.012$, $p=0.023$). Forcing the whole-artwork representation to be object-grounded trades away the composition and style cues that genre needs.

\begin{table}[t]
\caption{Genre macro-F1 on the SigLIP2 backbone under P1 with four object-evidence variants. $\Delta$ against the 0.896 baseline, paired-bootstrap $p$.}
\label{tab:bridge}
\begin{tabular}{lcc}
\toprule
Variant & macro-F1 & $\Delta$ ($p$) \\
\midrule
Baseline (class-conditioned mixture) & 0.896 & ref. \\
Object-evidence expert & 0.894 & $-0.002$ (0.69) \\
Object-guided pooling & 0.897 & $+0.001$ (0.87) \\
Evidence distillation & 0.883 & $-0.013$ (0.020) \\
Two-stage object pretraining & 0.884 & $-0.012$ (0.023) \\
\bottomrule
\end{tabular}
\end{table}

\begin{figure}[t]
\centering
\includegraphics[width=0.9\columnwidth]{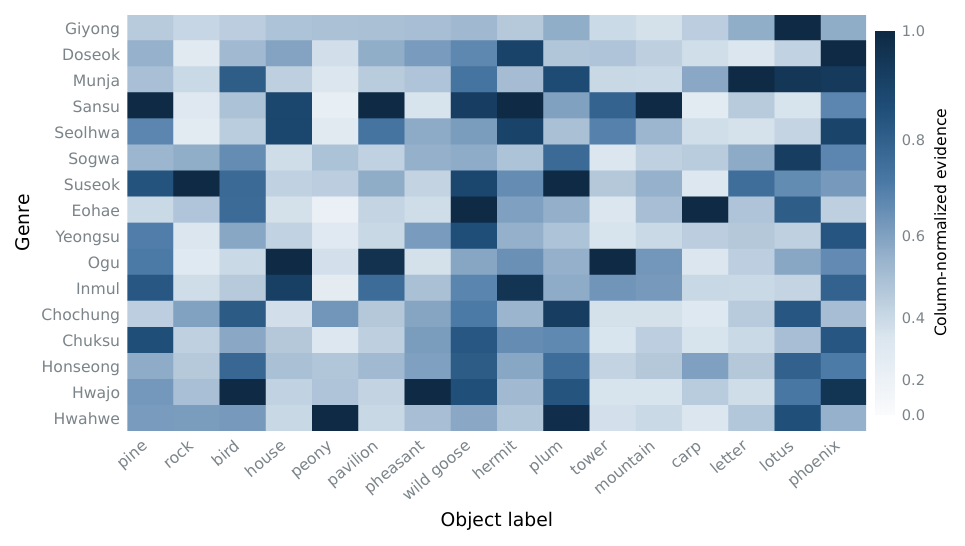}
\caption{Mean object evidence by genre, column-normalized over the sixteen most frequent objects.}
\label{fig:heatmap}
\end{figure}

Consistent with this, the model's residual confusions are object-structured. Across all 120 genre pairs the confusion rate correlates with the cosine similarity of per-genre object profiles, with a Spearman correlation of 0.49 (Mantel permutation test over genre labels, 20{,}000 permutations, $p<10^{-4}$, partial Spearman 0.42 after residualizing both the iconographic similarity and the confusion on the genre-pair size ratio). Because both quantities derive from the genre-labeled data, this is not independent evidence but a description of where errors fall, namely between genres that share symbolic objects such as composite and bird-and-flower paintings. Table~\ref{tab:sig} lists the most discriminative objects per genre by within-genre lift, which recover sensible data-driven associations, and Figure~\ref{fig:heatmap} shows the full per-genre profile. These associations are computed from the curatorial labels and are descriptive rather than a curated art-historical canon.

\begin{table}[t]
\caption{Data-driven object associations per genre by within-genre lift, for representative genres. These are descriptive statistics, not a curated canon.}
\label{tab:sig}
\begin{tabular}{lll}
\toprule
Genre & Top objects & Top motifs \\
\midrule
Character (munjado) & letter, filial-piety emblem & filial piety, virtue \\
Fish-and-crab (eohaedo) & carp, mullet, mandarin fish & crab, shrimp \\
Bird-and-flower (hwajodo) & pheasant, dove, duck & pheasant, dove \\
Daoist-figure (doseokhwa) & immortal, deer & longevity \\
Flower (hwahwedo) & peony, chrysanthemum, plum & wealth and glory \\
Landscape (sansuhwa) & boat, temple, pavilion & pavilion \\
Animal (chuksudo) & rabbit, deer, turtle, crane & deer, crane \\
\bottomrule
\end{tabular}
\end{table}

\subsection{Era is a collection confound}

Era behaves opposite to genre. The image is the strongest modality at 0.772 macro-F1 under P1 and 0.803 within a single collection under P2, while text reaches only 0.580 and fusion does not exceed the image, so era is carried by visual style that the captions do not capture. The within-collection era signal is genuine rather than a genre artifact, since predicting era from the genre label alone reaches only 0.460 within the dominant collection. The cross-source result, however, shows the signal is collection specific. A low-level analysis confirms the confound, since color saturation appears to decrease across eras over the mixed-source data but the trend nearly vanishes within a single collection. The apparent era style signature is therefore driven by the correlation between era and collection, not by time, which is why era does not transfer in Table~\ref{tab:cross}. A per-sample gate intended to suppress the redundant text for era does not learn to close, because the redundancy is a property of the label rather than of individual paintings. Further ablations on loss functions are in the supplementary material.

\section{Discussion}\label{sec:discussion}

The content-versus-style dichotomy has a direct consequence for digital cataloguing of minhwa. Genre is a content category that the described iconography recovers even after the stated label is masked, and it transfers across institutions, so caption-conditioned genre assignment is a dependable aid for metadata enrichment and consistency checking wherever a structured description already exists. Era is the opposite case. It is carried by collection-specific visual style rather than by time, and it collapses under cross-source transfer, so an era predictor trained on a mixed corpus reports an institutional signature rather than a chronological one. A curator should therefore treat automatic era attribution from a single aggregated collection as unreliable until it is validated across sources, whereas genre enrichment from captions is comparatively safe to deploy.

As a concrete example, consider the bird-and-flower panel (top-left) in Figure~\ref{fig:overlay}, a late-Joseon work from a held-out source institution. Its catalogue identifies the symbolic vocabulary as the crane, pine, peony, immortality herb, bamboo, and pigeon. The evidence map, computed without any region annotation, places its largest mass on a crane (top-patch score 0.999) and on the pine trunk (0.936), agreeing with the catalogue on those two elements. Where the catalogue names a peony, however, the evidence concentrates on plum blossoms (0.845) rather than on peony, a likely transcription effect since the description's symbolic list reads as the canonical bird-and-flower formula rather than a specific reading of this work. A cataloguer can use this overlay in two ways. First, to verify which catalogued symbols are visually grounded, since elements named in the description but absent from the evidence map are candidates for review. Second, to surface motifs the dominant-subject description does not mention, since the part-level detector was trained on a broader symbolic vocabulary. The evidence map is therefore an inspection layer that sits alongside the existing curatorial description rather than a substitute for the curatorial label.

The faithful-but-insufficient dissociation shapes how an object evidence map should be used in practice. The map is a usable inspection layer, since it localizes the symbolic regions that a patch-level genre predictor relies on and recovers sensible per-genre object profiles without any region annotation. It is not a statement of the deployed model's decision rule, and the object inventory it summarizes does not by itself name the genre. This cautions against reducing genre indexing to an object checklist, because a tradition whose categories turn on how peonies, tigers, and paired birds are arranged and rendered is not captured by a list of which symbols are present. Object grounding is best offered to curators as an explanation and retrieval aid that sits alongside the genre model rather than as a substitute for it.

The dataset's era-by-genre distribution also invites one art-historical dialogue. The vessels-and-implements still lifes concentrate in late Joseon, with 94 of the 149 such paintings dated to that period, the same period in which Kim's account of chungin patronage situates the flourishing of \emph{chaekgeori} and related still-life genres~\cite{ref:choson-objects}. The model-based reading and the art-historical reading agree on the where, while our confound analysis cautions that this concentration tracks the dominant institution's collecting history as well as the era, and a multi-collection aggregation would be expected to redistribute the apparent era signature across categories beyond vessels-and-implements.

More broadly, the study illustrates evaluation cautions that recur in heritage collections beyond minhwa. Single-institution dominance produces a style confound that is easily mistaken for a temporal or semantic signal, and within-dataset accuracy can overstate what a model has learned. Masking templated label text, splitting by source so that crops and whole paintings never cross the train and test boundary, and projecting a part-level detector onto whole artworks only out of fold are inexpensive practices that keep such confounds visible. We offer these, with the released protocols and weights, as transferable practice for computational study of symbol-dense pictorial traditions that fall outside the visual canon most pretrained vision models are calibrated on.

\section{Limitations}\label{sec:limitations}

Several limitations bound the scope of the claims. The study uses a single national collection with one curatorial sixteen-genre taxonomy, dominated by one institution, so the absolute numbers and the per-genre associations should be read within that scope, and what we expect to transfer is the methodology and the content-versus-style dichotomy rather than the values themselves. The raw data cannot be redistributed, so the work is not externally reproducible from images, and we release code, trained weights, fold assignments, seeds, and derived count matrices for partial verification rather than a runnable benchmark. The faithfulness evaluation is established for a patch-based surrogate genre predictor, not for the deployed cross-attention model, which consumes a single class token and cannot ingest a patch deletion, and we compare against random, patch-norm, and gradient-saliency orderings but do not run a retraining-based faithfulness check. The whole-artwork and part-level stages use different frozen backbones, and the grounding-does-not-help result is measured on the SigLIP2 backbone. The patch-level aggregation hyperparameters (LSE temperature 0.5, asymmetric-loss negative focus four and margin 0.05) were chosen by initial development without an exhaustive sweep, and the second-surrogate and spatial-IoU checks above are evidence the central finding is not narrowly tuned to these values, but a full hyperparameter grid is left to future work. An independent external corpus would extend the dichotomy prediction beyond the four labels in this dataset, which we did not run. The multimodal gain is contingent on a structured caption being present at prediction time, and the object and motif inventories are text-derived proxies rather than visually verified labels.

\section{Conclusion}\label{sec:conclusion}

MinhwaNet shows that multimodal fusion helps genre classification of Korean folk painting, with the gain concentrated on the long tail and robust to a within-collection control and to cross-source transfer, where cross-attention matches the transferable text signal that plain concatenation dilutes. A patch-level multiple-instance model significantly improves weakly-supervised recognition of localized symbolic objects, confirming that minhwa evidence is local. The central diagnostic finding is that object grounding is an interpretable bridge rather than a performance lever. The leakage-safe object evidence map is faithful to a patch-based surrogate genre predictor, verified by deletion and insertion against random, patch-norm, and gradient-saliency baselines, yet even a generous text-derived object inventory does not determine genre and an object-grounded objective does not raise accuracy, because genre is a relational property of how symbols are composed and rendered. A confound analysis separates a content label that transfers across institutions, genre, from a style label that does not, era. These results offer a multimodal model and a set of evaluation cautions for long-tailed cultural-heritage understanding of Korean folk painting, where modality complementarity, leakage-safe grounding, and collection confounds all matter. Future work includes spatial evaluation of object grounding against region annotations and prediction of symbolic meaning where text may carry knowledge the image does not.

\section*{Data and code availability}

The Korean folk painting collection used in this study was provided by the AI Hub platform (\url{https://aihub.or.kr}) operated by the National Information Society Agency (NIA), Republic of Korea, as dataset \#13 (Korean Traditional Folk Painting Production Data). Under the AI Hub usage policy the raw dataset, comprising images and annotations, cannot be redistributed and access is restricted to verified Korean nationals through the official portal. All code, trained model weights, fold assignments, seeds, and derived aggregate count matrices are released at \url{https://github.com/joonhyungbae/MinhwaNet}. Researchers wishing to replicate this work may apply for dataset access at \url{https://aihub.or.kr}.

\section*{Generative AI disclosure}

A generative AI assistant was used for language editing of author-written text. No analysis, experiment, figure, or numerical result was produced by an AI system, and the authors take full responsibility for the content of this article.

\bibliographystyle{ACM-Reference-Format}
\bibliography{sn-bibliography}

\end{document}